\documentclass[12pt]{article}

\begin{document}

\begin{center}
{\large \bf Limiting Case  of Modified Electroweak Model for Contracted Gauge Group}
\end{center}

\begin{center}
N.A.~Gromov \\
Department of Mathematics, Komi Science Center UrD, RAS, \\
Kommunisticheskaya st. 24, Syktyvkar 167982, Russia \\
E-mail: gromov@dm.komisc.ru
\end{center}

\begin{abstract}
The modification  of the Electroweak Model with  3-dimensional spherical geometry in the matter fields  space is suggested. 
 The  Lagrangian of this model is given by the sum of the {\it free} (without any potential term)  matter fields Lagrangian  and the standard gauge fields Lagrangian.
 The vector boson masses are  generated  by transformation of this
 Lagrangian from Cartesian coordinates to a   coordinates on the sphere $S_3$.
The limiting case of the bosonic part of the modified model, which corresponds to the contracted gauge group $SU(2;j)\times U(1)$ is discussed.
Within framework of the limit model Z-boson and electromagnetic fields can be regarded as  an external ones with respect to  W-bosons fields in the sence that  W-boson fields do not effect on these external fields. The masses of  all  particles of the Electroweak Model 
remain the same, but  field interactions in contracted model are more simple as compared with the standard Electroweak Model.
\end{abstract}



\section{Introduction }

The standard Electroweak Model  based on gauge group $ SU(2)\times U(1)$ gives a good  description of electroweak processes. The massive vector bosons predicted by the model was experimentally observed and have the masses $m_W=80GeV $ for charged W-boson and $m_Z=91GeV$ for neutral Z-boson.
At the same time the existence of the scalar field (Higgs boson)
has not been experimentally verified up to now. 
One expect that the future experiments on LHC will given definite answer.
The scalar field arise in the Standard Electroweak Model as a result of spontaneous symmetry
breaking by Higgs mechanism \cite{R-99}, which include three steps:
1) the potential of the  self-acting scalar field of the special 
form $V(\phi)=\mu^2\bar{\phi}\phi + \lambda (\bar{\phi}\phi)^2 $ is introduced {\it by hand} in the Lagrangian;
2) its minimal values are considered for imaginary mass $\mu^2 <0$ and are interpreted as degenerate vacuum;
3) one of the gauge equivalent vacuum is fixed and then all fields are regarded in the neighbourhood of this vacuum.

Sufficiently artificial Higgs mechanism  with its imaginary bare mass is a naive relativistic analog of the  phenomenological description of superconductivity \cite{Sh-09}.
Therefore there are a serious doubt 
whether electroweak symmetry is broken by such a Higgs mechanism, or by something else.
 The emergence  of  large number  Higgsless models  \cite{C-05}--\cite{MT-08} was stimulated by difficulties with Higgs boson. These models are mainly based on extra dimensions of different types or  larger gauge groups. 
A finite electroweak model without a Higgs particle which is used  a regularized quantum field theory \cite{E-66},\cite{E-67} was developed in \cite{MT-08}. 

The simple mechanism for generation of the vector boson masses in Electroweak Model was suggested in
\cite{Gr-07-R}--\cite{Gr-08}.
It is based on the fact that the quadratic form $\phi^{\dagger}\phi=\rho^2$ in the matter field space $\Phi_2$ is invariant with respect to gauge transformations. This quadratic form define the 3-dimensional sphere $S_3$ of the radius $\rho>0$ in the space $\Phi_2$ which is  $C_2$ or $R_4$ if real components are counted. 
The vector boson masses are  generated  by transformation of the {\it free}  (without any potential term like $V(\phi)$) Lagrangian from Cartesian coordinates to a   coordinates on the sphere $S_3$.
This transformation corresponds to transition from linear to nonlinear representation of the gauge group in the space of functions on $S_3$.
Higgs boson field does not appeared if the  sphere radius  does not depend on the space-time  coordinates $\rho=const$. 
For $\rho\neq const$ the real positive massless scalar field 
is presented  in the model  \cite{CFN-08},\cite{F-08}.
Such modified Electroweak Model keep all experimentally verified fields of the standard
Electroweak Model and does not include massive scalar Higgs boson, which is sequent of spontaneous symmetry breaking.
 The  concept of  generation masses for vector bosons in Electroweak Model by  transformation to radial coordinate is developed in \cite{MW-10}, \cite{ILM-10}, as well as in context of nonlinearly realized gauge groups in \cite{BFQ-08} or in context of nonlinear sigma models in \cite{ACLR-09}. 
 
One of the important ingredient of the Standard  Model is the simple group $ SU(2)$. More then fifty years  in physics it is well known 
the notion of group contraction \cite{IW-53}, i.e. limit operation, which transforms, for example, a simple or semisimple group to a non-semisimple one. From general point of view
for better understanding of a physical system it is  useful to  investigate  its properties for limiting values of their physical parameters.
In partucular, for a gauge model one of the similar limiting case corresponds to a  model with contracted gauge group.
The gauge theories for non-semisimple groups which Lie algebras admit invariant non-degenerate metrics was considered in \cite{NW-93},\cite{T-95}.

In the present paper we construct the analog of the bosonic part  of the modified  Electroweak Model 
as the gauge theory with the contracted non-semisimple group $SU(2;j)\times U(1)$. 
In Sec. 2, we recall the definition and properties of the contracted group $SU(2;j)$. 
In Sec. 3, we step by step modify
the main points of the Electroweak Model for the gauge group $SU(2;j)\times U(1)$ and
introduce the radial coordinates $R_{+}\times S_3$ in $R_4$ in the most convenient way \cite{F-08}. 
The use of the representation with the symmetric arranged contraction parameter \cite{Gr-94}
enables us to find transformation properties 
of gauge fields. After that the Lagrangian of the contracted model can be very easy obtained from the noncontracted  one  by the substitution (\ref{eq28}) or (\ref{eq29}).   
The limiting case of the modified Higgsless  Electroweak Model is regarded in Sec. 4.  
When contraction parameter tends to zero $j \rightarrow 0$ or takes nilpotent value $j=\iota$ the field space is fibered \cite{Gr-09} in such a way that
 electromagnetic and Z-boson fields are in the base whereas charged W-bosons fields are in the fiber.
The base fields can be interpreted as an external ones with respect to the fiber  fields.
In addition  field interactions are simplified under contraction.
Sec. 5 is devoted to the conclusions.

\section{ Special unitary group  $SU(2;j)$ and its contraction}

The contracted special unitary group $SU(2;j)$ is defined as a transformation group 
$$ 
\phi'(j)=
\left(\begin{array}{c}
\phi'_1 \\
j\phi'_2
\end{array} \right)
=\left(\begin{array}{cc}
	\alpha & j\beta   \\
-j\bar{\beta}	 & \bar{\alpha}
\end{array} \right)
\left(\begin{array}{c}
\phi_1 \\
j\phi_2
\end{array} \right)
=\Omega(j)\phi(j), \quad
$$
\begin{equation}
\det \Omega(j)=|\alpha|^2+j^2|\beta|^2=1, \quad  \Omega(j)\Omega^{\dagger}(j)=1
\label{g3}
\end{equation}  
of  the complex fibered vector space  $\Phi_2(j)$ which keep invariant the hermitian form 
\begin{equation}
\phi^\dagger(j)\phi(j)=|\phi_1|^2+ j^2|\phi_2|^2,
\label{g1}
\end{equation}  
where $\phi^\dagger(j)=(\bar{\phi_1},j\bar{\phi_2}), $ bar denotes  complex conjugation,
parameter $j=1, \iota$ and $\iota$ is nilpotent unit $\iota^2=0.$  For nilpotent unit the following heuristic rules be fulfiled: 
1) division of a real or complex numbers by $\iota$ is not defined, i.e. for a real or complex $a$ the expression $\frac{a}{\iota}$
is defined only for $a=0$, 
2) however identical nilpotent units can be cancelled $\frac{\iota}{\iota}=1.$

 Contracted group and fibered space are obtained for nilpotent value of the parameter $j= \iota$,  whereas unitary group $SU(2)$ and  usual complex vector  space correspond to  $j=1.$ The space $\Phi_2(\iota)$ is fibered with one dimensional base $\left\{\phi_1\right\}$ and one dimensional fiber $\left\{\phi_2\right\}$ \cite{Gr-94},\cite{Gr-09}. 
 It has two hermitian forms: first  in the base $\bar{\phi_1}\phi_1=|\phi_1|^2$ and second in the fiber  $\bar{\phi_2}\phi_2=|\phi_2|^2$.  Both forms can be unified in  one formula (\ref{g1}).

The  generators of the corresponding Lie algebra $su(2;j)$
$$    
  T_1(j)= j\frac{i}{2}\left(\begin{array}{cc}
	0 & 1 \\
	1 & 0
\end{array} \right)=j\frac{i}{2}\tau_1, \quad 
T_2(j)= j\frac{i}{2}\left(\begin{array}{cc}
	0 & -i \\
	i & 0
\end{array} \right)=j\frac{i}{2}\tau_2, 
$$
\begin{equation} 
T_3(j)= \frac{i}{2}\left(\begin{array}{cc}
	1 & 0 \\
	0 & -1
\end{array} \right)=\frac{i}{2}\tau_3, 
\label{g7}
\end{equation} 
with $\tau_k$ being  Pauli matrices, 
are subject of commutation relations
$$  
[T_1(j),T_2(j)]=-j^2T_3(j), \quad [T_3(j),T_1(j)]=-T_2(j), 
$$
\begin{equation} 
 [T_2(j),T_3(j)]=-T_1(j).
\label{g8}
\end{equation}
  The general element of $su(2;j)$ is given by
\begin{equation} 
 T(j)=\sum_{k=1}^{3}a_kT_k(j)= \frac{i}{2}\left(\begin{array}{cc}
	a_3 & j(a_1-ia_2) \\
	j(a_1+ia_2) & -a_3
\end{array} \right)=-T^{\dagger}(j).
\label{g8-1}
\end{equation}
 
There are two more or less equivalent way of group contraction. We can put the contraction parameter equal to the nilpotent unit $j=\iota$  or  tend it to zero $j\rightarrow 0$.  Sometimes it is  convenient to use the first (mathematical) approach, sometimes the second (physical) one. For example, the matrix $\Omega(j)$  (\ref{g3}) has non-zero nilpotent non-diagonal elements for $j=\iota$,
whereas for $j\rightarrow 0$ they are formally equal to zero. Nevertheless both approaches lead to the same final results.
 
 For $j=\iota$ it follows from (\ref{g3}) that 
$\det \Omega(\iota)=|\alpha|^2=1,$ i.e. $\alpha=e^{i\varphi},$ therefore
\begin{equation} 
\Omega(\iota)= 
\left(\begin{array}{cc}
e^{i\varphi}	 & \iota\beta   \\
-\iota\bar{\beta}	 & e^{-i\varphi}
\end{array} \right) \in SU(2;\iota), \quad   
\beta=\beta_1+i\beta_2 \in {\bf C}. 
\label{g9}
\end{equation}
The simple group $SU(2)$ is contracted to the non-semisimple  one $SU(2;\iota)$, which is isomorphic to the real  Euclid group $E(2).$
First two generators of the Lie algebra $su(2;\iota)$ are commute 
and the rest commutators are given by (\ref{g8}). 
 
The unitary group $U(1)$ and the electromagnetic subgroup $U(1)_{em}$ are essential ingredients 
 of the  Electroweak Model
Their actions in the   fibered  space $\Phi_2(\iota)$ are given by the same matrices as in the complex space $\Phi_2$, namely
\begin{equation}
u(\beta)=e^{\beta Y}=\left(\begin{array}{cc}
	e^{i\frac{\beta}{2}} & 0 \\
0	 & e^{i\frac{\beta}{2}}
\end{array} \right), \quad
u_{em}(\gamma)=e^{\gamma Q}=\left(\begin{array}{cc}
	e^{i\gamma} & 0 \\
0	 & 1
\end{array} \right),
\label{g13}
\end{equation}  
where $Y=\frac{i}{2}{\bf 1}, \; Q=Y+T_3.$

Representations of groups $SU(2;\iota), U(1), U(1)_{em}$   are linear ones,  that is they are realised  by linear operators in  the   fibered  space $\Phi_2(\iota)$.

\section{ Electroweak Model for $SU(2;j)\times U(1)$ gauge group}

The fibered space $\Phi_2(j)$  can be obtained from $\Phi_2$ by 
substitution $\phi_2 \rightarrow j\phi_2$, 
which induces another ones for Lie algebra $su(2)$ generators
$T_1 \rightarrow jT_1,\; T_2 \rightarrow jT_2,\;T_3 \rightarrow T_3. $
As far as the gauge fields take their values in Lie algebra, we can substitute gauge fields instead of transformation of generators, namely
\begin{equation}
A_{\mu}^1 \rightarrow jA_{\mu}^1, \;\; A_{\mu}^2 \rightarrow jA_{\mu}^2,\; \;A_{\mu}^3 \rightarrow A_{\mu}^3, \;\;
B_{\mu} \rightarrow B_{\mu}.
\label{eq28}
\end{equation}  
These substitutions in the Electroweak Model based on $SU(2)\times U(1)$ gauge group  result in the limiting case of  Electroweak Model based on contracted  gauge group $SU(2;\iota)\times U(1)$, when contraction parameter take the nilpotent value $j=\iota$.

The  bosonic Lagrangian of the contracted Electroweak Model is given by the sum
\begin{equation}
L(j)=L_A(j) + L_{\phi}(j),
\label{eq1}
\end{equation}
where
$$  
 L_A(j)=\frac{1}{2g^2}\mbox{tr}(F_{\mu\nu}(j))^2 + \frac{1}{2g'^2}\mbox{tr}(\hat{B}_{\mu\nu})^2= 
 $$
 \begin{equation}
=  -\frac{1}{4}[j^2(F_{\mu\nu}^1)^2+j^2(F_{\mu\nu}^2)^2+(F_{\mu\nu}^3)^2]-\frac{1}{4}(B_{\mu\nu})^2
\label{eq2}
\end{equation}
is the gauge fields Lagrangian for $SU(2;j)\times U(1)$ group and
\begin{equation}   
  L_{\phi}(j)= \frac{1}{2}(D_\mu \phi(j))^{\dagger}D_\mu \phi(j)            
\label{eq3}
\end{equation}  
is the {\it free} (without any potential term) matter field Lagrangian (summation on the repeating Greek indexes is always understood).  
Here $D_{\mu}$ are the covariant derivatives
 \begin{equation}
D_\mu\phi(j)=\partial_\mu\phi(j) + g\left(\sum_{k=1}^{3}T_k(j)A^k_\mu \right)\phi(j) + g'YB_\mu\phi(j),
\label{eq4}
\end{equation} 
where $T_k(j)$ are given by (\ref{g7})
 and 
$Y=\frac{i}{2}{\bf 1}$ is generator of $U(1).$ 
Their actions on components of $\phi(j)$ are given by
$$
D_\mu \phi_1=\partial_\mu \phi_1 + \frac{i}{2}(gA_\mu^3+g'B_\mu)\phi_1 + j^2\frac{ig}{2}(A_\mu^1-iA_\mu^2)\phi_2,
$$
\begin{equation}
D_\mu \phi_2=\partial_\mu \phi_2 - \frac{i}{2}(gA_\mu^3-g'B_\mu)\phi_2 + \frac{ig}{2}(A_\mu^1+iA_\mu^2)\phi_1.
\label{eq5}
\end{equation}

The gauge fields 
$$ 
A_\mu (x;j)=g\sum_{k=1}^{3}T_k(j)A^k_\mu (x)=g\frac{i}{2}\left(\begin{array}{cc}
	A^3_\mu  & j(A^1_\mu -iA^2_\mu ) \\
j(A^1_\mu + iA^2_\mu ) & -A^3_\mu 
\end{array} \right),
$$
\begin{equation}
 \hat{B}_\mu (x)=g'YB_\mu (x)=g'\frac{i}{2}\left(\begin{array}{cc}
	B_{\mu} & 0 \\
0 & B_{\mu}
\end{array} \right)
\label{eq6}
\end{equation}
 take their values in Lie algebras $su(2;j),$  $u(1)$ respectively,  and the stress tensors are
$$ 
F_{\mu\nu}(x;j)={\cal F}_{\mu\nu}(x;j)+[A_\mu(x;j),A_\nu(x;j)]=
$$
$$
=g\frac{i}{2}\left(\begin{array}{cc}
	F^3_\mu  & j(F^1_\mu -iF^2_\mu ) \\
j(F^1_\mu + iF^2_\mu ) & -F^3_\mu 
\end{array} \right), 
$$
\begin{equation} 
B_{\mu\nu}=\partial_{\mu}B_{\nu}-\partial_{\nu}B_{\mu},         
\label{eq7} 
\end{equation}
or in components 
$$
F_{\mu\nu}^1={\cal F}_{\mu\nu}^1  + g(A_\mu^2A_\nu^3-A_\mu^3A_\nu^2), \quad
F_{\mu\nu}^2={\cal F}_{\mu\nu}^2 +g(A_\mu^3A_\nu^1-A_\mu^1A_\nu^3),
$$
\begin{equation}
F_{\mu\nu}^3={\cal F}_{\mu\nu}^3 + j^2g(A_\mu^1A_\nu^2-A_\mu^2A_\nu^1),
\label{eq8} 
\end{equation}
where ${\cal F}_{\mu\nu}^k=\partial_\mu A_\nu^k-  \partial_\nu A_\mu^k. $

For  $\Omega(j)\in SU(2;j),\; e^{i\omega}\in U(1)$ the gauge transformations of the fields are as follows
$$
\phi^{\Omega}(j)=\Omega(j)\phi(j), \quad \phi^{\omega}(j)=e^{i\omega}\phi(j),
$$
$$
A_\mu^{\Omega}(x;j)= \Omega(j) A_\mu(x;j)\Omega^{-1}(j) -\partial_{\mu}\Omega(j)\cdot \Omega(j)^{-1}, \quad A_\mu^{\omega}(x;j)= A_\mu(x;j),
$$
\begin{equation}
B_{\mu}^{\Omega}= B_{\mu}, \quad B_{\mu}^{\omega}= B_{\mu}- 2\partial_{\mu}\omega
\label{eq9}
\end{equation} 
and the Lagrangian $L(j)$ (\ref{eq1}) is invariant under $SU(2;j)\times U(1)$ gauge group.

The Lagrangian   $L(j)$ (\ref{eq1}) describe  massless fields. 
In a standard approach to generate   mass terms for the vector bosons the "`sombrero"' potential is added to the matter field Lagrangian 
$  L_{\phi}(j=1)$ (\ref{eq3}) and after that the Higgs mechanism is used.
The different way was recently suggested \cite{Gr-07-R}--\cite{Gr-08}  and is based on the fact that the quadratic form $\phi^{\dagger}\phi=\rho^2$
is invariant with respect to gauge transformations. This quadratic form define the 3-dimensional sphere $S_3$ of the radius $\rho>0$ in the target space $\Phi_2$ which is  $C_2$ or $R_4$ if real components are counted. In  other words  the radial coordinates $R_{+}\times S_3$ are introduced in $R_4.$ 
The vector boson masses are easy generated by the  transformation of Lagrangian $L(j=1)$ (\ref{eq1}) to the   coordinates on the sphere $S_3$. 
Higgs boson field does not appeared if the  sphere radius  does not depend on the space-time  coordinates $\rho=R=const$ \cite{Gr-07-R}--\cite{Gr-08}. For $\rho\neq const$ the real positive massless scalar field --- analogy of dilaton or kind of Goldstone mode --- is presented  in the model  \cite{F-08}.

Let us introduce the radial coordinates   in the most convenient way following  \cite{F-08}. Write $\phi(j)$ as
\begin{equation}
\phi(j)=\left(\begin{array}{c}
 \phi_1\\
j\phi_2
\end{array} \right)=\rho\left(\begin{array}{c}
 \chi_1\\
j\chi_2
\end{array} \right)=\rho\left(\begin{array}{cc}
 \chi_1 & -j\bar{\chi}_2\\
j\chi_2 & \bar{\chi}_1
\end{array} \right) 
\left(\begin{array}{c}
 1\\
0
\end{array} \right) \equiv \rho h(j) \varphi_0,
\label{eq10}
\end{equation} 
where $\rho $ is a positive function,
then from $\phi^{\dagger}(j)\phi(j)=\rho^2 $ it follows that
\begin{equation}
\chi^{\dagger}(j)\chi(j)=\bar{\chi_1}\chi_1 + j^2 \bar{\chi_2}\chi_2=1
\label{eq11}
\end{equation} 
and the matrix $h(j)$ is unimodular $\det h(j)=1$ and unitary $h^{\dagger}(j)h(j)=1.$   
So the vector $\chi(j)= h(j) \varphi_0 \in S_3(j) $ or the matrix $h(j)\in SU(2;j)$ defines the coordinates on the sphere $S_3(j)$ (\ref{eq11}).
Radial variable is invariant $\rho^{\Omega}=\rho,\; \rho^{\omega}=\rho, $
 but the matrix $h(j)$ transforms  under the gauge transformations as
\begin{equation}
h^{\Omega}(j)=\Omega(j)h(j),\quad
h^{\omega}(j)=\left(\begin{array}{cc}
 e^{i\omega}\chi_1 & -j e^{-i\omega}\bar{\chi}_2\\
j e^{i\omega}\chi_2 &  e^{-i\omega}\bar{\chi}_1
\end{array} \right)=h(j) e^{i\omega \tau_3}, 
\label{eq12}
\end{equation} 
therefore covariant derivatives   of $\rho$ and $h(j)$ 
assume the forms
\begin{equation}
D_{\mu}\rho=\partial_{\mu}\rho, \quad
D_{\mu}h(j)=\partial_{\mu}h(j) + A_{\mu}(x;j)h(j) +  \hat{B}_{\mu}(x)h(j)\tau_3.
\label{eq13}
\end{equation} 
Using (\ref{eq10}), (\ref{eq13}) we obtain 
$$ 
D_{\mu}\phi(j)=\partial_{\mu}\rho h(j) +\rho D_{\mu}h(j)\varphi_0= 
$$
$$
=h(j)\left\{\partial_{\mu}\rho +  \rho  \left[h^{\dagger}(j)\partial_{\mu}h(j) + h^{\dagger}(j)A_{\mu}(x;j)h(j) +  \hat{B}_{\mu}(x)\tau_3 \right]\right\}\varphi_0= 
$$
\begin{equation}
=h(j)\left\{\partial_{\mu}\rho +  \rho  \left[W_{\mu}(x;j) +  \hat{B}_{\mu}(x)\tau_3 \right]\right\}\varphi_0, 
\label{eq14}
\end{equation} 
where the new vector field is introduced
\begin{equation}
W_{\mu}(x;j)= h^{\dagger}(j)A_{\mu}(x;j)h(j) + h^{\dagger}(j)\partial_{\mu}h(j)
\label{eq15}
\end{equation} 
with components
\begin{equation}
W_{\mu}(x;j)=\frac{i}{2}g\left[j\left(W_{\mu}^1\tau_1 + W_{\mu}^2\tau_2 \right) + W_{\mu}^3\tau_3 \right].
\label{eq16}
\end{equation} 
As a result the matter field Lagrangian (\ref{eq3}) takes the form
\begin{equation}   
  L_{\phi}(j)= \frac{1}{2}\left(\partial_{\mu}\rho\right)^2 +  
  \frac{1}{2} \frac{\rho^2}{4} \left(\sqrt{g^2+g'^2}\right)^2 \left(Z_{\mu}\right)^2 +
 j^2 \frac{\rho^2g^2}{4}W_{\mu}^{+}W_{\mu}^{-,}
\label{eq17}
\end{equation} 
where the new fields are introduced
\begin{equation}   
W_{\mu}^{\pm}=\frac{1}{\sqrt{2}}\left(W_{\mu}^1 \mp iW_{\mu}^2\right), \quad 
Z_{\mu}=\frac{gW_{\mu}^3 + g'B_{\mu}}{\sqrt{g^2+g'^2}}, \quad 
A_{\mu}=\frac{g'W_{\mu}^3 - gB_{\mu}}{\sqrt{g^2+g'^2}}.
\label{eq18}
\end{equation} 
These fields  are invariant under the gauge transformations of $SU(2;j)$: 
$ X^{\Omega}=X,\;\; X=W_{\mu}^{\pm}, W_{\mu}^{3}, Z_{\mu}, A_{\mu} $
and are transforms under those  of $U(1)$ as 
$$    
\left(W_{\mu}^{\pm}\right)^{\omega}=e^{\mp2i\omega}W_{\mu}^{\pm}, \quad 
\left(W_{\mu}^3\right)^{\omega}= W_{\mu}^3 +\frac{2}{g}\partial_{\mu}\omega, 
$$
\begin{equation}
Z_{\mu}^{\omega}=Z_{\mu}, \quad 
A_{\mu}^{\omega}=A_{\mu} + \frac{2}{e}\partial_{\mu}\omega, \quad 
e=\frac{gg'}{\sqrt{g^2+g'^2}}.
\label{eq19}
\end{equation} 
It follows from (\ref{eq28})  that substitutions of these fields are
\begin{equation}
W_{\mu}^{\pm} \rightarrow jW_{\mu}^{\pm}, \;\; Z_{\mu} \rightarrow Z_{\mu},\; \;
A_{\mu} \rightarrow A_{\mu}.
\label{eq29}
\end{equation}

It is easy to check, that the stress tensors of the fields $A_{\mu}(x;j)$ and $W_{\mu}(x;j)$ (\ref{eq15}) are connected by 
\begin{equation}
W_{\mu\nu}(x;j)=h^{\dagger}(j)F_{\mu\nu}(x;j)h(j), 
\label{eq20}
\end{equation} 
therefore
\begin{equation}
\frac{1}{2}\mbox{tr}(F_{\mu\nu}(x;j))^2=\frac{1}{2}\mbox{tr}(W_{\mu\nu}(x;j))^2
\label{eq21}
\end{equation} 
and the gauge fields Lagrangian (\ref{eq2}) can be written as
\begin{equation}
L_A(j)=-j^2\frac{1}{4}\left[(W_{\mu\nu}^1)^2 + (W_{\mu\nu}^2)^2 \right] -\frac{1}{4}\left[(W_{\mu\nu}^3)^2 + (B_{\mu\nu})^2 \right].
\label{eq22}
\end{equation} 
The first two terms assume the form
$$ 
-j^2\frac{1}{2}\left(\nabla_{\mu}W_{\nu}^{+} - \nabla_{\nu}W_{\mu}^{+} \right)\left(\nabla_{\mu}W_{\nu}^{-} - \nabla_{\nu}W_{\mu}^{-} \right)=
$$
\begin{equation}
=j^2\left\{ -\frac{1}{2}{\cal{W}}^{+}_{\mu\nu}{\cal{W}}^{-}_{\mu\nu} + \frac{igP - g^2S}{2\sqrt{g^2+g'^2}} \right\},
\label{eq23}
\end{equation} 
where
$$ 
\nabla_{\mu}W_{\nu}^{\pm}=(\partial_{\mu} \mp ig W_{\mu}^{3})W_{\nu}^{\pm},\quad
{\cal W}^{\pm}_{\mu\nu}=\partial_{\mu}W^{\pm}_{\nu} - \partial_{\nu}W^{\pm}_{\mu},
$$ 
$$
P={\cal{W}}^{+}_{\mu\nu}\left[W_{\mu}^{-}(gZ_{\nu}+g'A_{\nu}) - W_{\nu}^{-}(gZ_{\mu}+g'A_{\mu}) \right] - 
$$
$$
 - {\cal{W}}^{-}_{\mu\nu}\left[W_{\mu}^{+}(gZ_{\nu}+g'A_{\nu}) - W_{\nu}^{+}(gZ_{\mu}+g'A_{\mu}) \right],
$$
$$ 
S=\left[W_{\mu}^{+}(gZ_{\nu}+g'A_{\nu}) - W_{\nu}^{+}(gZ_{\mu}+g'A_{\mu}) \right] \times
$$
\begin{equation}
\times 
\left[W_{\mu}^{-}(gZ_{\nu}+g'A_{\nu}) - W_{\nu}^{-}(gZ_{\mu}+g'A_{\mu}) \right].
\label{eq24}
\end{equation}
The last two terms in (\ref{eq22}) are 
$$ 
-\frac{1}{4}\left[(W_{\mu\nu}^3)^2 + (B_{\mu\nu})^2 \right]=
-\frac{1}{4}\left[( {\cal{W}}^{3}_{\mu\nu} + j^2H_{\mu\nu})^2 + (B_{\mu\nu})^2 \right]=
$$
$$ 
= -\frac{1}{4}\left[({\cal{W}}^{3}_{\mu\nu})^2 + (B_{\mu\nu})^2 + 2j^2{\cal W}_{\mu\nu}^3H_{\mu\nu} + j^4(H_{\mu\nu})^2 \right]=
$$
\begin{equation}
= -\frac{1}{4}\left[({\cal{Z}}_{\mu\nu})^2 + ({\cal A}_{\mu\nu})^2 + 2j^2{\cal W}_{\mu\nu}^3H_{\mu\nu} + j^4(H_{\mu\nu})^2 \right],
\label{eq25}
\end{equation} 
where
$$
W^3_{\mu\nu}={\cal W}^{3}_{\mu\nu} + H_{\mu\nu}, \quad 
H_{\mu\nu}= -ig\left(W_{\mu}^{+}W_{\nu}^{-} - W_{\mu}^{-}W_{\nu}^{+} \right),
$$
\begin{equation}
{\cal W}^{3}_{\mu\nu}=\partial_{\mu}W^3_{\nu} - \partial_{\nu}W^3_{\mu}, \quad
{\cal Z}_{\mu\nu}=\partial_{\mu}Z_{\nu} - \partial_{\nu}Z_{\mu}, \quad
{\cal A}_{\mu\nu}=\partial_{\mu}A_{\nu} - \partial_{\nu}A_{\mu}.
\label{eq26}
\end{equation} 
The bosonic Lagrangian (\ref{eq1}) assumes the form
$$
L(j)= \frac{1}{2}\left(\partial_{\mu}\rho\right)^2  -\frac{1}{4} ({\cal A}_{\mu\nu})^2 -\frac{1}{4}({\cal{Z}}_{\mu\nu})^2 
 + \frac{1}{2} \frac{\rho^2}{4} \left(\sqrt{g^2+g'^2}\right)^2 \left(Z_{\mu}\right)^2 +
$$
$$
+ j^2 \left\{-\frac{1}{2}{\cal{W}}^{+}_{\mu\nu}{\cal{W}}^{-}_{\mu\nu}  +\frac{\rho^2g^2}{4}W_{\mu}^{+}W_{\mu}^{-}
+ \frac{igP - g^2S - (g{\cal Z}_{\mu\nu} + g'{\cal A}_{\mu\nu})H_{\mu\nu} }{2\sqrt{g^2+g'^2}}\right\} -
$$
\begin{equation}
-j^4 \frac{1}{4}(H_{\mu\nu})^2 = L_b + j^2L_f + j^4L_h.
\label{eq27}
\end{equation}

For the  gauge group $SU(2;j)\times U(1)$
the quadratic form $\phi^{\dagger}(j)\phi(j)=\rho^2$
is invariant with respect to the gauge transformations and  defines the  sphere $S_3(j)$   
in the target space $\Phi_2(j)$. 
As it was mentioned the vector boson masses are automatically (without any Higgs mechanism) generated by the  transformation of the {\it free} Lagrangian (\ref{eq1})  to the Lagrangian (\ref{eq27}) expressed in some coordinates on the sphere $S_3(j)$. 
And this is true for both values of the contraction parameter $j=1$  and $j=\iota$. 
If the  sphere radius does not depend on the space-time  coordinates  $\rho=R=const$, 
 then  $\partial_{\mu}\rho =0 $ and the real positive scalar field $\rho$ as well as Higgs boson field are not present in the model   \cite{Gr-07-R}--\cite{Gr-08}. 
In this case  Lagrangian (\ref{eq27})  
 describes massless  vector fields $A_{\mu}, m_A=0$ (photon),  massive vector field $Z_{\mu}$ with the mass $m_Z=\frac{1}{2}R \sqrt{g^2+g'^2}$  (Z-boson), massive vector fields $W_{\mu}^{\pm}$ with identical mass $m_W=\frac{1}{2}gR$ (W-bosons)  and interactions of these fields. $W$- and $Z$-bosons have been    observed and have the masses   $  m_W=80 GeV,$ $ m_Z=91 GeV $,
so   only experimentally verified  fields of Electroweak Model are included in Lagrangian.

\section{Limiting case  of modified  Electroweak Model}

We assume $\rho=R=const,\; \partial_{\mu}\rho =0 $ in this section.
Let contraction parameter tends to zero $j^2\rightarrow 0$, then the contribution $L_f$  of W-bosons fields to the Lagrangian (\ref{eq27}) will be small in comparison  with the contribution $L_b$ of Z-boson and electromagnetic fields. 
The  term $L_h$ being fourth order in $j$ can be neglected.
In other words the limit Lagrangian $L_b$ includes    Z-boson and electromagnetic fields and charded W-bosons fields does not effect on these fields.
The  part $L_f$ form a new Lagrangian for W-bosons fields and their interactions with other fields. 
The appearance of two Lagrangians $L_b$ and $L_f$ for the limit model is in correspondence with two hermitian forms of fibered  space $\Phi_2(\iota)$, which are invariant under the action of  contracted gauge group $SU(2;\iota)$. 
 Electromagnetic  and Z-boson fields can be regarded  as external ones with respect to the W-bosons fields. 

In mathematical language   the  field space $\left\{ A_{\mu}, Z_{\mu}, W_{\mu}^{\pm}\right\}$
is fibered after contraction $j=\iota$ to the base $\left\{ A_{\mu}, Z_{\mu}\right\}$ and the fiber 
$\left\{W_{\mu}^{\pm}\right\}.$ 
(In order to avoid terminological misunderstanding let us stress that we  have in view locally trivial fibering, which
is defined by the projection $pr:\; \left\{ A_{\mu}, Z_{\mu}, W_{\mu}^{\pm}\right\} \rightarrow \left\{ A_{\mu}, Z_{\mu}\right\}$ in the field space. 
This fibering is understood in the context of  semi-Riemannian geometry \cite{Gr-09} and has nothing to do with the principal fiber bundle.)
Then $L_b$ in (\ref{eq27}) presents Lagrangian in the base and $L_f$ is Lagrangian in the fiber.  In general, properties of a fiber are depend on a points of a base and not the contrary.
In this sense fields in the base can be interpreted as  external ones with respect to fields in the fiber. 

Let us note that field interactions in contracted model are more simple as compared with the standard Electroweak Model 
due to nullification of some terms.
For example, the last term $j^4L_h$ in (\ref{eq27}) disappears as having fourth order in $j\rightarrow 0$.

\section{Conclusions} 

We have discussed the limiting case of the modified Higgsless Electroweak Model \cite{Gr-07-R}--\cite{Gr-08}, which corresponds to the contracted gauge group $SU(2;j)\times U(1)$, where $j=\iota$ or $j \rightarrow 0$.
The masses of the all experimentally verified particles involved in the Electroweak Model remain the same under contraction, but interactions of the fields are changed in two aspects. 
Firstly  all field interactions become more simpler due to nullification of some terms in Lagrangian.   
Secondly  interrelation  of the fields become more complicated. All fields are divided on two classes: fields in the base
(Z-boson and electromagnetic) and fields in the fiber (W-bosons). 
The base  fields  can be interpreted as  external ones with respect to the fiber fields, i.e.  Z-boson and electromagnetic fields can interact with W-bosons fields, but  W-bosons fields do not effect on these fields within framework of the limit model.
Let us note that the fibering of the field space under contraction in the modified Electroweak Model
is very similar to those in the standard Electroweak Model \cite{Gr-10}. The only exception is the presence  of the scalar Higgs boson field in the base in the last case.

The author is grateful to V.V. Kuratov for helpful discussions.
This work has been supported in part by the Russian Foundation for Basic Research, grant 08-01-90010-Bel-a
and the program "`Fundamental problems of nonlinear dynamics"' of Russian Academy of Sciences.


\end{document}